# Developments for the IPNS Data Acquisition System Upgrade

J.P. Hammonds*, R.R. Porter*, A. Chatterjee*, T.G. Worlton*, P.F. Peterson*,
J.W. Weizeorick[§], P.M. De Lurgio[§], I. Naday[§]
Argonne National Laboratory
Intense Pulse Neutron Source Division*
Electronics and Computing Technology Division[§]

D.J. Mikkelson, R.L. Mikkelson
University of Wisconsin Stout
Department of Mathematics, Statistics and Computer Science

## Abstract

The Intense Pulsed Neutron Source has been an operating user facility for more than 20 years. Development of an upgrade for the data acquisition system has been in progress for some time now. Now that the initial installation on the test bed instrument[1] (HRMECS) is completed new features and possibilities are starting to come forward. Run setup is moving from large monolithic programs to more flexible scripting. EPICS[2,3,4] based controls allow flexible and extendable instrument configurations. Hardware is also being made more flexible. This flexibility allows implementation of new methods of binning data, that better match instrument resolution. This allows better use of memory and disk space. Increased flexibility has also allowed use of the new time-of-flight modules in applications past their original intent.

## Introduction

### A Need for New Data Acquisition at the IPNS Facility

The Intense Pulsed Neutron Source (IPNS) is an accelerator driven spallation neutron source that has operated as a user facility at Argonne National Laboratory since May of 1981. It provides slow neutrons for neutron scattering studies of condensed matter. At IPNS, 13 instruments on 12 beam lines operate independently to provide neutrons for a wide range of experiments. Detectors on these instruments include:

- Single element $^3$He gas filled tubes
- $^3$He gas filled linear position sensitive detectors (LPSD's)
- $^3$He gas filled multi-wire detectors
- Scintillator based Anger cameras

In order to facilitate maintenance and upgrade ability, these detectors must use a minimal complement of hardware that is unique to a particular system and must use as much commercially available hardware as possible.

The base system that in use at IPNS has been in operation since the start of operation in 1981[5,6,7,8] and has had minimal upgrades over the last 2 decades. At present, the old hardware is starting to systematically fail, and components that were originally used are no longer commercially available. At the same time, instruments enhancements have added detectors that have consumed hardware spares.

Software maintenance and planning for this system is also problematic. In the past, most of the software design focused around having more limited resources and are generally tied to a hardware interface between the VAX/PDP instrument computers. In the new system, hardware interfaces are replaced with software interfaces. The EPICS control system that was chosen for control of the data acquisition system (DAS) uses TCP/UDP protocols to allow communication between hardware and the user. This eliminates the tie between acquisition hardware and choice of proprietary interface to a user computer. Also, programs that are composed of smaller, easy to manage Java classes replace large monolithic codes for setup, display and manipulation.

A program to develop and replace the aging hardware, and provide a path for instrument enhancement has been underway for some time now. Installation on the first instrument is now complete and the next round of installations is underway.

### Requirements of the Data Acquisition System Upgrade

Before starting on an upgrade of our present system, a list of requirements was compiled. From our old system we had the requirements:

- Collect time of flight data from any of the detector types listed above.
- Time range must extend to at least 66ms. This accommodates collection times that extend into a second 30Hz

pulse of the accelerator. This allows for study of longer wavelength neutrons.
- Time resolution <= 1/8 microsecond.
- Must be able to handle at least $2^{16}$ detector elements.
- User must be able to easily reconfigure time-binning scheme.
- Be able to deal with at least 16 million individual time bins.
- Must provide interface to automate instrument control.

In addition to matching the performance of the old system, a list of new requirements was developed
- Make it possible to use any combination of detectors on an instrument.
- Extend the number of pixels available on an instrument.
- Make the chosen architecture independent of the choice of proprietary computer hardware.
- Improve instrument diagnostics.
- Provide variable channel width capabilities to all types of detectors.

## System Description

Detector Data Acquisition System Hardware

As described above, this system was intended to be designed around as much commercially available hardware as possible. A schematic of the system is shown in Figure 1. The base of this system is a C sized VXI crate that contains:
- A commercial VME processor board, running the vxWorks real time operating system
- A custom built readout control module (ROC)
- A custom built time-of-flight (TOF) module.

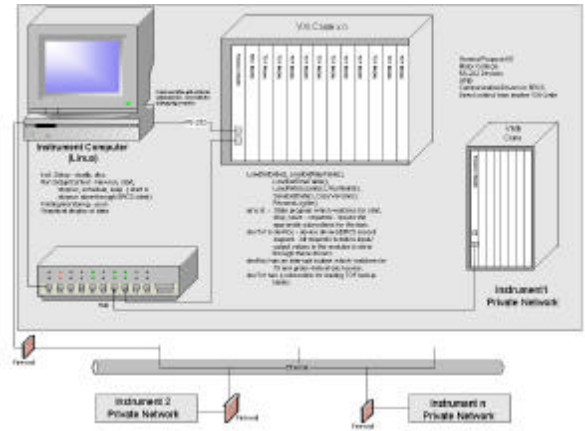

**Figure 1. Structure of DAS System**

The processor board provides the system control interface. In this module:
- EPICS databases and device drivers are loaded here to communicate with hardware.
- The EPICS Channel Access Server runs to provide an interface with the user.
- A state program runs to coordinate start/stop queues
- Interrupt routines run to process and histogram data events
- A data sender periodically sends data to the instrument computer for live data access.

We have been able to transition through three different processor boards with a minimum of effort, as the needs of developments have required more resources. We are currently developing around the Motorola MVME2304 PowerPC board with 128MB of RAM.

The ROC module provides input of common system signals such as an accelerator trigger pulse and a veto signal. It also provides test input signals that can be passed into the system for hardware diagnostics. These signals can also be used in more complicated systems as a trigger to coordinate acquisition of more complex detectors.

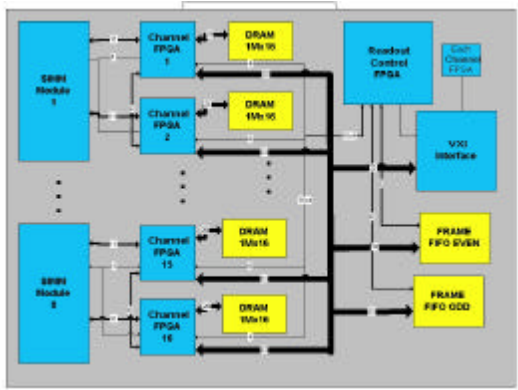

**Figure 2. Time of Flight Module Block Diagram**

The TOF module (Figure 2) is the real heart of the system. This module was originally conceived for use with single element tube detectors. The design seemed flexible enough that it was built to also handle input from LPSDs. The module accepts 16 differential inputs, which are routed through 8 changeable analog input modules. This allows the module to be used with 16 single element tubes or 8 LPSDs. Since 2 of the inputs are routed to the analog input module, handshaking between the two ends of an LPSD can be handled on this module. Each of these analog input modules has a corresponding pair of field programmable gate arrays (FPGAs). Each FPGA interfaces to a memory chip that can provide table lookup capabilities. As a pulse (or pair of pulses) are detected by the analog input module, a trigger to the corresponding channel FPGA(s) is raised.

In the case of a single ended tube, this trigger locks in a time stamp. This time stamp is the number of 10MHz clock cycles from the last accelerator pulse. This latched time is then used as an address to the channel lookup table, which provides a bin number for this event. This allows the creation of a minimum time binning resolution of 100ns. Since the channel lookup table is a 1MB table, the maximum time allowable is 104ms, or 3 times the accelerator pulse period. This time stamp is combined with a detector ID, which is unique in this VXI crate, to form a data event word. This event is stored as a 32 bit word that has 16 bits of position and 16 bits of time. The event word is stored in a board level ping-pong FIFO buffer, which is emptied and placed in a histogram in memory by the crate processor board.

Similarly, the pulse from an LPSD triggers two FPGAs. One of these FPGAs will perform the time lookup procedure described above while the second performs a lookup for the position using pulse height data from both sides of the detector. This information is then combined as output from one of the FPGAs into a data word as described above.

The design of this board has proven adaptable. Because the FPGAs can be programmed, they provide a mechanism to collect data from a neutron anger camera. For these detectors, three of these boards are used to collect data from 36 photo multiplier tubes (PMT) that view a scintillator plate. Each of the three TOF modules collect pulse height information from 12 of the PMTs plus an additional data channel that represents a trigger from the sum of all PMTs. The 13$^{th}$ channel is used to trigger a time lookup as described for the single element detector. The trigger signal comes from an input on the ROC module and is fed to all 39 channels on the three TOF modules. This trigger notifies the TOFs when to make the pulse height and time measurements. The data from each channel is gathered into the board FIFO buffer. The data is read by the crate processor, is processed through a position extraction algorithm, and then histogrammed into the same memory histogram space as the other detectors.

In general limits imposed by the system described above are:
- 176 detectors or 88 LPSDs/crate (11 slots)
- 16 bits of position allow 65K pixels per crate
- 16 bits of time allow 65K time channels per detector
- 128MB on the system board leaves about 100MB for histogram memory. With 4 byte words, that gives about 25M channels per system crate.

Data Acquisition Software
The control for this system is based on the Experiment Physics and Industrial Control System (EPICS). EPICS is a control package that has been developed by a community of more than a hundred facilities world wide. It is a good base for starting a project of this nature and allows for building on the work of others. This choice was also driven by the co-location of the Advanced Photon Source, one of the major

EPICS developers, and the Electronics & Computing Technology Division, that has been developing systems for the APS, on site at Argonne.

EPICS uses a logical database structure to provide an interface between hardware and software. Device drivers for modules are tied to record support interfaces. References to the record support are added to database definitions that describe I/O points in the system. Each "record" in the database is given a name, which must be unique to the overall system. A communication library is also provided which allows communication with I/O points via a TCP/UDP/IP interface. This allows communication between a user interface computer and the hardware, as well as soft records that help control the system.

A state notation language is also provided with EPICS to allow change of state event handling. A state notation program has been developed which watches for start, stop and save requests. On start, the program loads parameters from a run file to set up hardware parameters. This loads detector discrimination levels, histogram memory addressing, and time lookup tables, as well as setting up other instrument controls. On stop, processing of events stops and data is saved.

We experienced one problem using EPICS for a data acquisition system. Transfers between the instrument crate and user computers using the supplied Channel Access protocol were limited to 16KB per transfer. Data spectra were converted to floats by one of the common display tools and the overall process for this kind of transfer was a bit cumbersome. To overcome this problem, a new LiveDataServer[9] Java class was developed as part of our ISAW[10,11] data visualization/manipulation package. This class accepts spectra via a UDP stream. The stream contained enough information to allow the server to gather enough information from a run file. ISAW clients could then access the server via a TCP/IP request. This allows not only live data display, but remote live display of data.

Instrument Setup Software, a Journey
Historically, IPNS has used a file-based system to control parameters for a particular experiment. In the old VAX based system, there was one program for setting up run files for all instruments. This provided a common interface for all instruments. The program maintained enough flexibility to provide setup for different types of experiments. This program was a large monolithic beast. It was also so heavily tied to VAX/VMS FORTRAN that it was impossible to migrate to another system. At the time this effort began, Java had emerged at IPNS. An attempt to reproduce the function of the original FORTRAN code in Java was made but this was determined to be unmanageable. One piece of code that survived this effort was the RunfileBuilder class that extended the Runfile (reader) class. This class had enough functionality to store and then write all of the information that was needed in the run file. Most of the methods needed to populate the information needed in the file were in place. All that remained was a good wrapper around this class. As the decision to scrap a monolithic program was coming close, a scripting language, written for a package (ISAW) to read, display and analyze data was emerging at IPNS. This package was written in Java, provided a mechanism for wrapping the RunfileBuilder methods and was going to be in use for instrument display and analysis. The effort to write the wrappers, develop scripts, and test this out took only a couple of weeks and provided a much quicker method to add new functionality to the system. Before the switch, only single ended tubes were supported and no integrated control of sample environment was included. Within a couple of months of the switch, LPSDs and environment control methods were added and functioning on an instrument. Within one year the capability for area detectors is in place and support for new sample environment is available.

Run setup evolved around the existing concept of a run file that contained all information on instrument parameters, controlled devices, data organization and the data itself. This file is read by a run start procedure and facilitates loading instrument hardware with appropriate parameters. Problems of data organization within the file had to be resolved to accommodate some of the new requirements of the system. Problems fixed inside of the run file structure include:

- Limits on the number of pixels used. In the old structure, this was limited, except in the case of an area detector, to 65K pixels. In the new structure the limit has been changed to accommodate 65K pixels in one crate and up to 2 billion total time channels. It is still possible to change the crate limit if the number of time channels can be reduced.
- Limits on detector combinations. Complexities in the file structure made it impossible to use LPSDs and area detectors in the same instrument. It also limited the number of single element detectors that could be used with LPSDs. In the new system, all detectors are treated in the same way. All are considered multi-pixel detectors and information is all stored in the descriptor tables.

Variable Width Time Channels
In the past, data from most detectors used at IPNS was collected as a series of time of flight histograms with bins that had a constant Δt channel width. Unfortunately, this type of binning does not closely match the resolution of an instrument. On a diffractometer, for instance, a scheme where the ratio of channel width and time are kept constant better represents the instrument resolution. With this system, it is fairly simple to change the time map for the histogram by changing the values in the channel lookup table. This allows data to be collected which makes better use of instrument resolution and system memory. An example of this principle is shown in Figures 3, 4 and 5. Each image shows data from three runs of equal duration. The first shows the entire extent of the data while the second and third show detail at short and long times respectively. The three runs used the following binnings:
- Constant dt = 3.3μs.
- Constant dt/t = 0.0007.
- Constant dt/t = 0.0005.

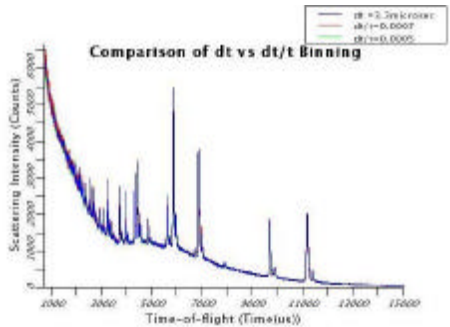

**Figure 3. Comparison of constant dt/t with constant dt binning**

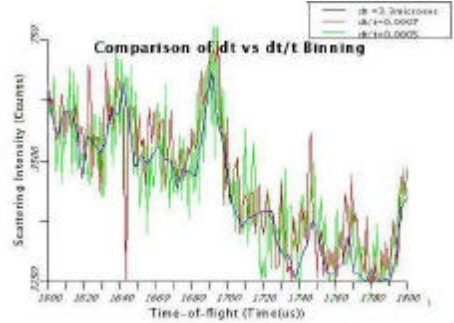

**Figure 4. Comparison data at short times** - Constant dt/t binning shows better peak resolution than constant dt binning.

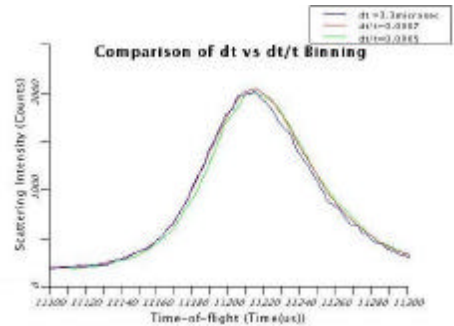

**Figure 5. Comparison data at long times** - Constant dt/t and constant dt binning are roughly equivalent.

The total number of channels for dt/t binning can be given by $n = (1/C) \ln (t_{max}/t_{min})$. $T_{min} = 700\mu s$ and $T_{max} = 24000\mu s$. The number of channels for each spectrum is therefore 7060, 5049 and 7069 respectively. Therefore, the data quality can be made much better with less than or equal time channels for constant dt/t binning.

Instrumentation Control
Unfortunately, the task of upgrading a system in this manner was not limited to changing the DAS and a few programs on a computer. A larger share of the work will come in supplying support for sample environment controls. This equipment includes temperature controls, motor controls, vacuum readout, and pressure cells. This also includes a category of equipment that will see a more rapid turnover of equipment. This area is where the use of EPICS will pay off. Other laboratories have completed much of the groundwork in this area. By leveraging on the work of others, we have been able to pick up drivers and EPICS databases for most of the equipment that we will need. All that was needed was a method to integrate this into the overall control system for our experiments, and to define a scheme that makes adding a new device transparent. To approach this we developed a two-layer approach.

Step 1. First, look for EPICS support of our existing equipment, if this is appropriate. If this is not available we weigh the options of replacing the equipment with supported hardware or writing a device driver. Next, make the chosen device communicate with EPICS, following a defined naming convention for EPICS database records that helps facilitate system integration.

Step 2. Add records to the EPICS system that determine if the device is ready to acquire data. If not, a veto record should flag the DAS to wait. Next, a text file is defined which describes control points of the device, their associated database record names, and a default value. A sample file is shown here:

```
deviceName=Displex
deviceNameDbSignal=deviceName
controllerName=Lakeshore330
dbDevice=lkshr330
vetoSignal=vetoRec
ancIoc=A
#User Parameters
numUserParameters=3
UserParameterName1=SetPoint
UserParameter1=35.0
UserParameterDbSignal1=setpt
UserParameterName2=Minimum
UserParameter2=40.0
UserParameterDbSignal2=tmin
UserParameterName3=Maximum
UserParameter3=60.0
UserParameterDbSignal3=tmax
#Inst Parameters
numInstParameters=1
InstParameterName1=HeaterPower
InstParameter1=1.0
InstParameterDbSignal1=htr_pwr
InstParameterOptions1=OFF,Low,Medium,
High
```

This file provides space via user parameters for specific user input such as temperature set point and range, and instrument specific parameters such as heater settings, PID values, etc.

The setup file is interpreted by a method in a run setup script. The user would select from a list of appropriate devices and input control parameters. This information will then be written into the run file that is then loaded as part of the run start procedure.

## Conclusions

A system that has proven to be flexible is now in place at IPNS. The hardware that has been designed is adaptable enough to extend well beyond the original intent. Modular programming is a must. Each component should be designed to be as independent as possible in order to allow for testing and change.

Now that the base of the system is in place, we are coming up with new ways to extend it. One holdback to a full-scale replacement of the system in place is money. We are now starting to think of ways to produce hybrid systems, which allow new and old to coexist in a unified manner. This should allow us to get the most mileage out of both systems.

## Acknowledgements

Work performed at ANL is supported by the U.S. DOE-BES under contracts No. W-31-109-ENG-38. In particular, we would like to thank Chun-Keung Loong, Alexander Kolesnikov, Joe Fieramosca, Dong Feng Chen and Ken Takeuchi who suffered patiently through the debugging process while this system was installed on the HRMECS instrument. We would also like to thank members of the technical staff at IPNS and ECT at Argonne National Laboratory for assistance in the assembly, testing and installation of the system components. Of special note are Chris Piatak and Tom Walsh who have helped manage most of the effort for this installation.